\documentclass{cimento}

\usepackage{graphicx}
\usepackage{amssymb}
\usepackage{amsmath}
\usepackage{color}

\newcommand{\cosths}{\ensuremath{\cos\,\theta^{*}}}

\title{Search for Single Top Production at HERA}
\author{G.~Brandt\from{ins:x}\ETC\thanks{on behalf of the H1 and ZEUS collaborations.}}
\instlist{\inst{ins:x} 
Deutsches Elektronen-Synchrotron, Notkestr.~85, 22607 Hamburg, Germany}
\PACSes{\PACSit{13.85.t,~13.90.i,~14.65.Ha}{}}
\begin{document}

\maketitle

\begin{abstract}

A search for single top production in $ep$ collisions 
using the complete high energy data from HERA is presented.
This search is based on the analysis of events containing isolated leptons
(electrons or muons) and missing transverse momentum~$P_{T}^{\mbox{miss}}$. 
In the absence of a signal, an upper limit on the top
production cross section $\sigma_{ep\rightarrow etX} < 0.16$~pb is established
at the $95\%$ confidence level, corresponding to an upper bound on the
anomalous magnetic coupling $\kappa_{tu\gamma} < 0.14$.
The search is complemented by a search for events containing an isolated tau lepton 
and $P_{T}^{\mbox{miss}}$ and the measurement of $W$ boson polarisation fractions.

\end{abstract}

\section{Introduction}
\label{sec:intro}

The HERA $ep$ collider, located at DESY in Hamburg, Germany,
was in operation in the years 1992--2007. Protons with
an energy up to $920$~GeV were brought into collision with electrons
\footnote{unless otherwise stated the term electrons also refers to positrons
in the following.}
of energy $27.6$~GeV at two experiments, H1 and ZEUS, each of which collected
about $0.5$~fb$^{-1}$ of data.
Together with measuring the structure of the proton, the deep inelastic scattering
(DIS) at HERA provided an ideal
environment to study rare processes and search for new particles and physics beyond 
the Standard Model (BSM).
In particular, the centre-of-mass energy up to $\sqrt s = 320$~GeV makes the production of
single top quarks possible. However a cross section of 
$\mathcal{O}(1~\mbox{fb})$ in the Standard Model (SM) is too small to
measure this process~\cite{smtop}. Any observation of single top
quark events in the HERA data would therefore be a clear sign of new physics
beyond the SM.

\begin{figure}[tp]
(a) \includegraphics[width=.45\textwidth]{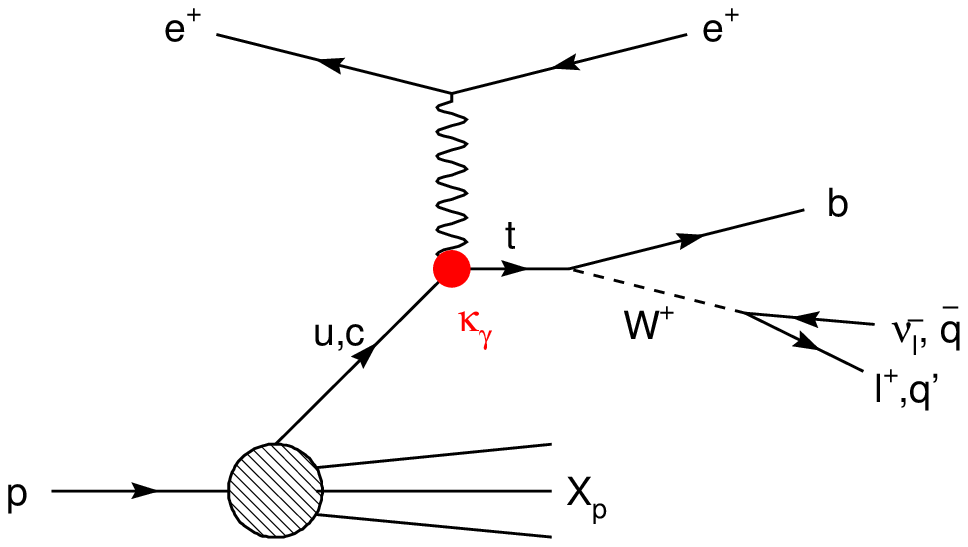}
(b) \includegraphics[width=.45\textwidth]{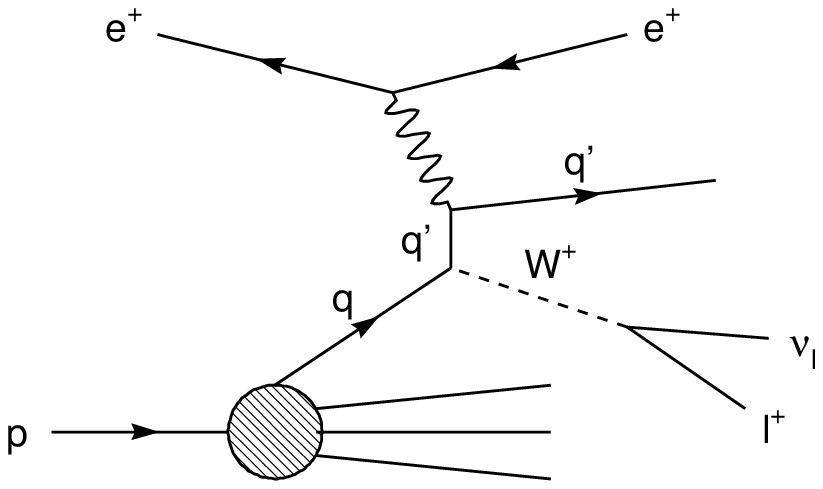}
\caption{(a) Diagram of
anomalous single top proudction in an $ep$ collision via an effective FCNC coupling
$\kappa_{ t u \gamma}$.
(b) Diagram of single $W$ production in the SM,
the process $ep \rightarrow eW^{\pm}$($\hookrightarrow l\nu$)$X$, 
which is the main SM background to single top production. }
\label{fig:feyn}
\end{figure}

In several extensions of the SM the top quark is predicted to undergo
Flavour Changing Neutral Current (FCNC) interactions, which could lead to a sizeable
anomalous single top production cross section at HERA~\cite{toptheory}.
Any such process is described by an effective Lagrangian where the interaction 
of a top quark with $u$-type quarks via a photon is described by a magnetic coupling $\kappa_{tU\gamma}$,
as illustrated in fig.~\ref{fig:feyn}(a).
The ANOTOP MC generator is used to model such events~\cite{anotop}.
The main SM contribution to the signal topology is the production
of real $W$ bosons via photoproduction with subsequent leptonic decay
$ep\rightarrow eW^{\pm}$($\hookrightarrow l\nu$)$X$, as illustrated in
fig.~\ref{fig:feyn}(b).
This process is modelled using the event generator EPVEC~\cite{epvec}.
For both processes the signature is an event containing an isolated lepton and
missing energy from the neutrino. From single top decays $t\rightarrow bW$ a 
prominent hadronic system due to the $b$-jet is expected, while for SM $W$ production
the hadronic system $X$ has typically low transverse momentum $P_{T}^{X}$.

\section{Events with Isolated Leptons and $P_{T}^{\rm miss}$}
\label{sec:isointro}

\begin{figure}[tp]
(a) \includegraphics[width=.45\textwidth]{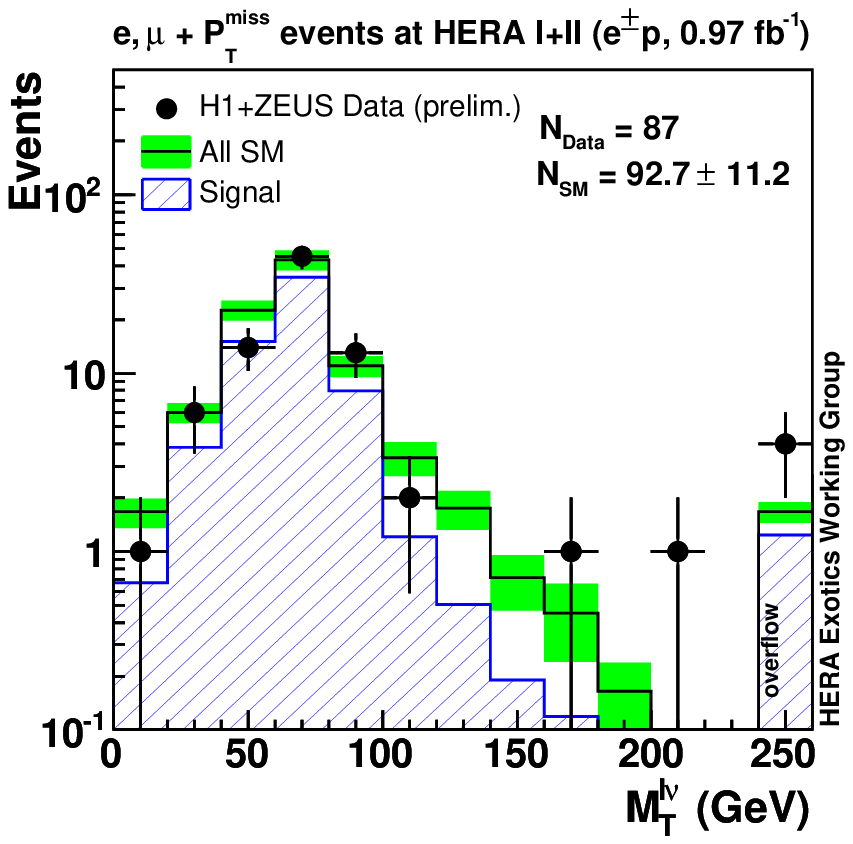}
(b) \includegraphics[width=.45\textwidth]{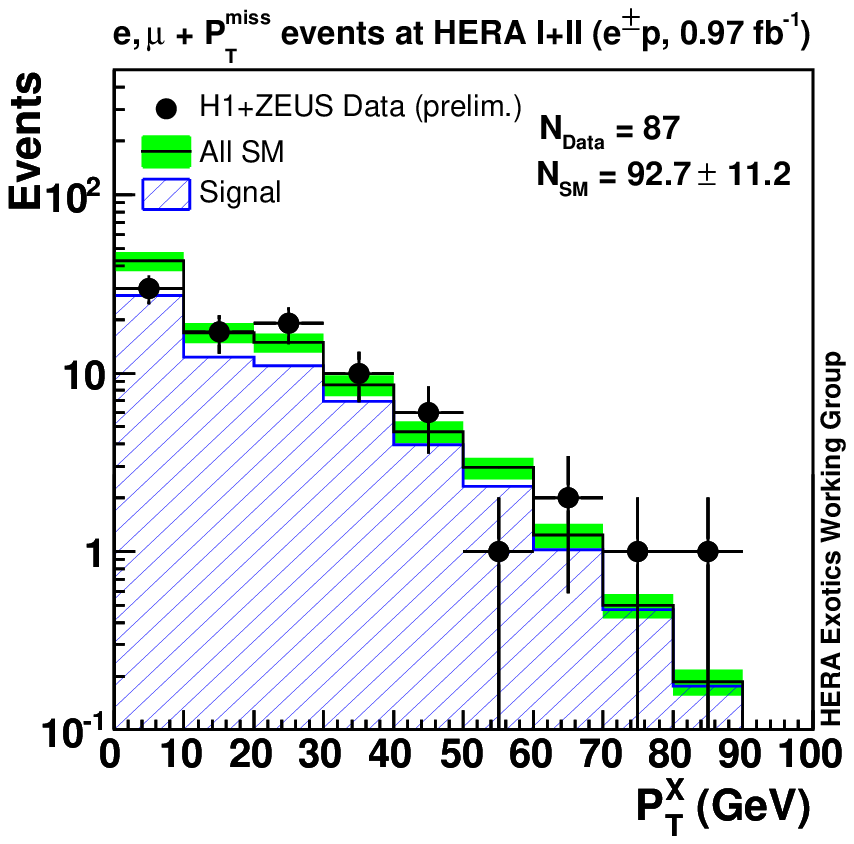}
  \caption{Distribution of events with isolated leptons in 
	the combined H1+ZEUS $e^{\pm}p$ HERA~I+II data.
	Shown is the transverse mass $M_{T}^{l\nu}$ (a) and hadronic
	transverse momentum $P_{T}^{X}$ (b).	
	The data (points) are compared to the
	SM expectation (open histogram). The signal ($W$ production) component of the SM
	expectation is given by the striped histogram. $\rm N_{Data}$ is the
	total number of data events observed and $\rm N_{SM}$ is the total SM
	expectation. The total uncertainty on the SM expectation is given by the
	shaded band.}
\label{fig:h1zeus}
\end{figure}

The search for single top quarks at HERA is based on 
events containing a high $P_{T} > 10$~GeV 
isolated electron or muon and missing~$P_{T} > 12$~GeV. 
Such events have been observed at
HERA~\cite{isoleph1origwpaper,isoleph1newwpaper,isolepzeusorigwpaper,zeustop}.
An excess of HERA~I data events (1994--2000) % which is mostly in $e^{+}p$ collisions)
compared to the SM prediction at large hadronic transverse
momentum $P_{T}^{X}$ was reported by the H1 Collaboration~\cite{isoleph1newwpaper}.
The significance of this excess did not increase with the inclusion of the HERA~II 
data~\cite{h1isolepnew} and was not confirmed by the ZEUS Collaboration~\cite{zeusisolepnew}.
Both experiments have combined their data and
good overall agreement with the SM is observed~\cite{h1andzeusisolepnew}.
Figure \ref{fig:h1zeus} shows the distribution of the transverse mass $M_{T}^{l\nu}$ and the $P_{T}^{X}$ of these events for the combined H1+ZEUS sample. The SM prediction is dominated
by $W$ production.

Background to SM $W$ and anomalous single top production enters the electron 
channel due to mismeasured neutral
current events and the muon channel due to muon pair 
events in which one muon escapes detection, both cases resulting in
apparent missing transverse momentum.
Charged current background, which contains intrinsic missing transverse momentum,
enters in both lepton channels, where a hadronic final
state particle is misidentified as an isolated electron or muon.

\section{Search for Anomalous Single Top Quark Production}

\begin{figure}[ht]
  \begin{center}
    \includegraphics[width=.63\textwidth]{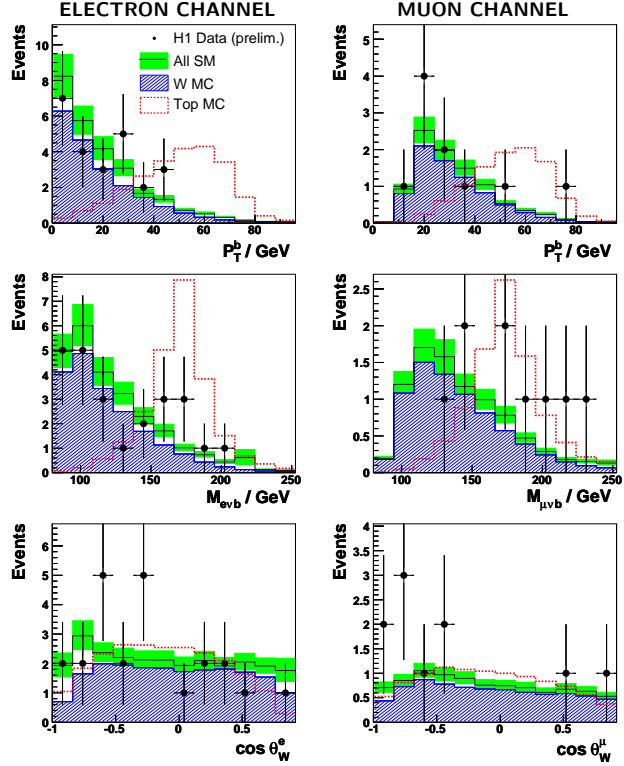}
  \end{center}
  \caption{
Control distributions of observables used to differentiate
anomalous single top production from SM background processes in the
top preselection. Shown are the transverse momentum of the reconstructed $b$ quark $P_T^b$,
the reconstructed top mass $M_{\ell\nu b}$ and the $W$ decay angle $\cos\theta_W^\ell$
for the electron channel (left column) and muon channel (right column). The data are shown as
points, the total SM expectation as open histogram with errors as shaded band.
The $W$ production component used to train the multi-variate
discriminators is shown as hatched histogram and the shape of the top signal MC is shown
as dashed line. }
  \label{fig:toppresel}
\end{figure}

The H1 search for single top production is based on the H1 sample of events with
isolated leptons selected in the HERA~I+II data, corresponding to a luminosity of $482$~pb$^{-1}$~\cite{h1isolepnew}. It presents an update of a previous top search
in HERA~I data~\cite{h1top}.
The first step in the analysis forms a top preselection in this event sample, by demanding
good top quark reconstruction and lepton charge compatible with single top production~\cite{h1topnew}.
Three observables that have been found to be suitable for separating SM $W$ and top production are
investigated in this preselection. These are the transverse momentum of the
reconstructed $b$ quark candidate $P_{T}^{b}$, the reconstructed top mass $M_{\ell\nu b}$, 
and the $W$ decay angle $\cos\theta_W^\ell$ calculated
as the angle between the lepton momentum in the $W$ rest frame and the $W$ direction in
the top quark rest frame.
The observed data distributions of these quantities, shown in fig.~\ref{fig:toppresel}, agree 
well with the SM expectation within the uncertainties. No evidence for single top production is observed.

%%%

\begin{figure}[tp]
  \includegraphics[width=.94\textwidth]{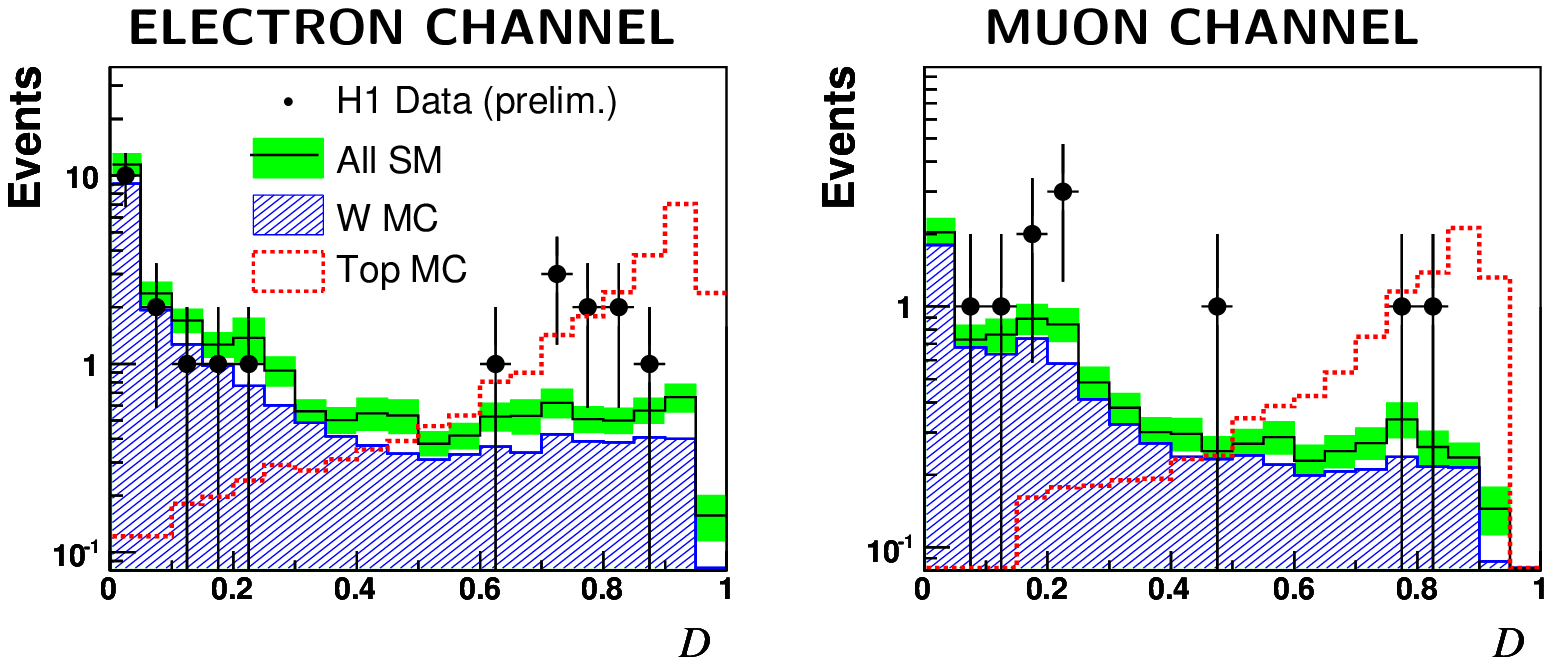} 
\caption{Discriminator output distributions for the electron channel (left) 
and muon channel (right) from a phase space density estimator using a
range search algorithm. The data are shown as
points, and the total SM expectation as open histogram with systematical and statistical errors
added in quadrature as shaded band. The $W$ production component used to train the multi-variate
discriminators is shown as hatched histogram and the shape of the top signal MC is shown
as dashed line. }
  \label{fig:topdiscr}
\end{figure}

The observables are then combined into a multivariate discriminator, which is trained using
ANOTOP as the signal model and EPVEC as the background model. The discriminator is based on 
a phase space density estimator using a range search algorithm~\cite{tmva}.
The resulting discriminator output distributions for the electron and muon channels, 
shown in fig.~\ref{fig:topdiscr}, are found to provide
good separation between $W$ and top MC events.

\begin{figure}[tp]
  \begin{center}	
    \includegraphics[width=.705\textwidth]{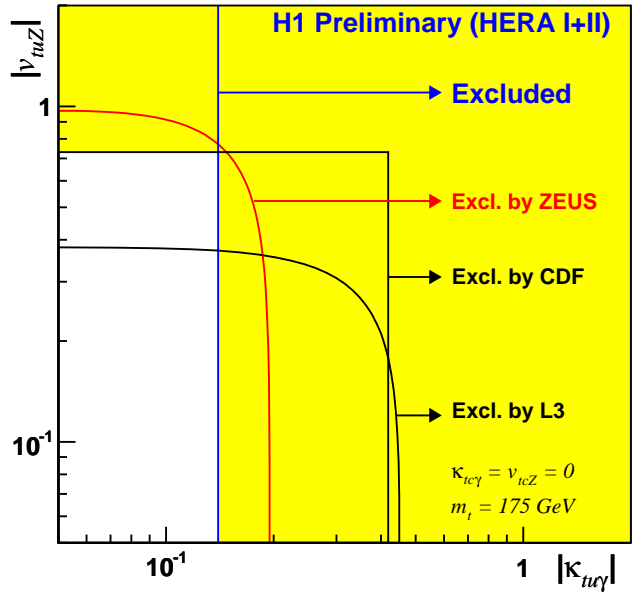}
  \end{center}
  \caption{Exclusion limits at 95\% CL on  the anomalous $\kappa_{tu\gamma}$
	and $v_{tuZ}$ couplings obtained at HERA (H1~\cite{h1topnew} and
	ZEUS~\cite{zeustop} experiments), LEP (L3 experiment~\cite{l3top})
	and at the TeVatron (CDF experiment~\cite{cdftop} Run I). Anomalous couplings
	of the charm quark $\kappa_{tc\gamma}$ are neglected.
	Limits are shown assuming a top mass $m_t=175$~GeV.}
  \label{fig:toplimit}
\end{figure}

\begin{figure}[tp]
(a) \includegraphics[width=.45\textwidth]{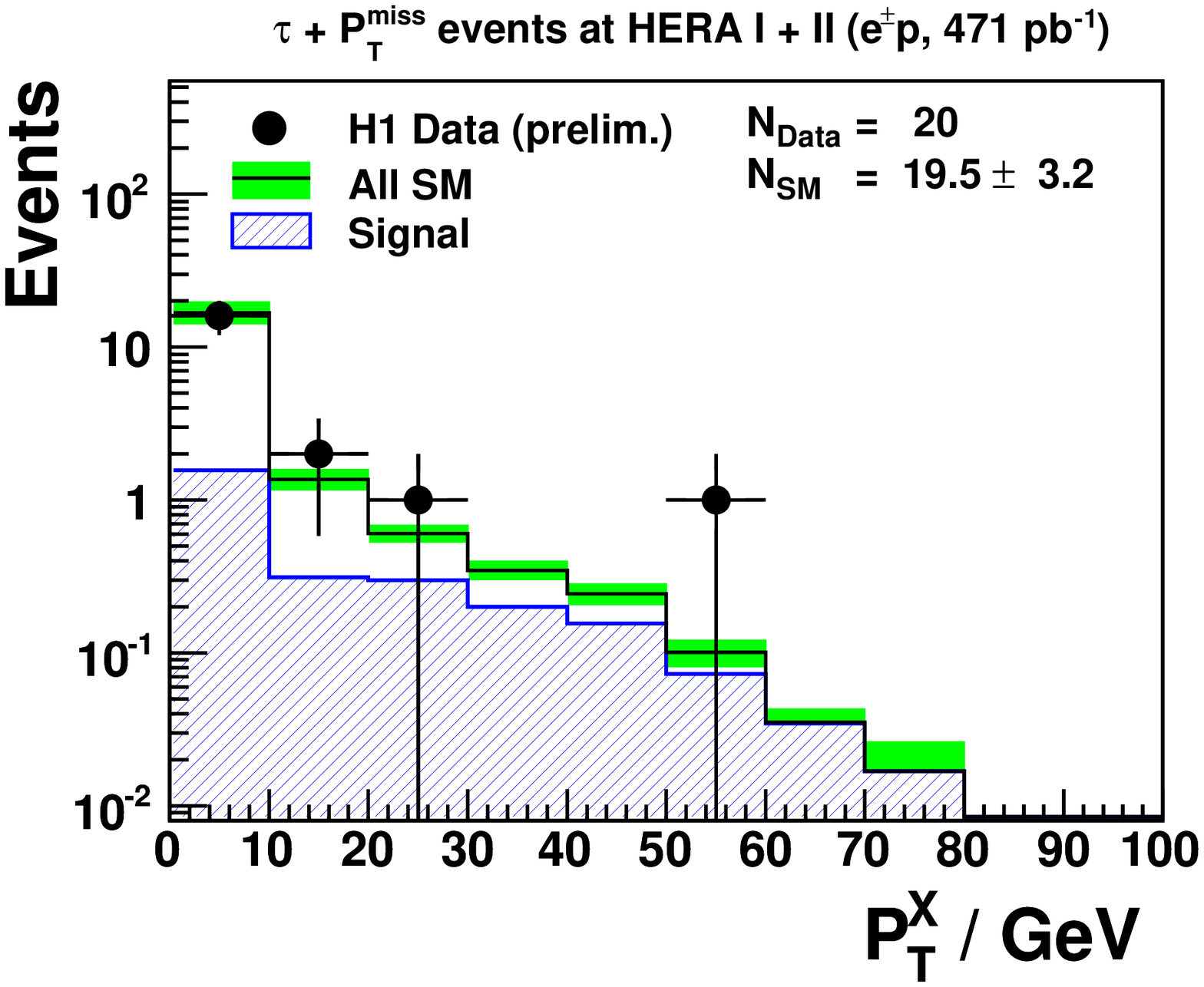}
(b) \includegraphics[width=.45\textwidth]{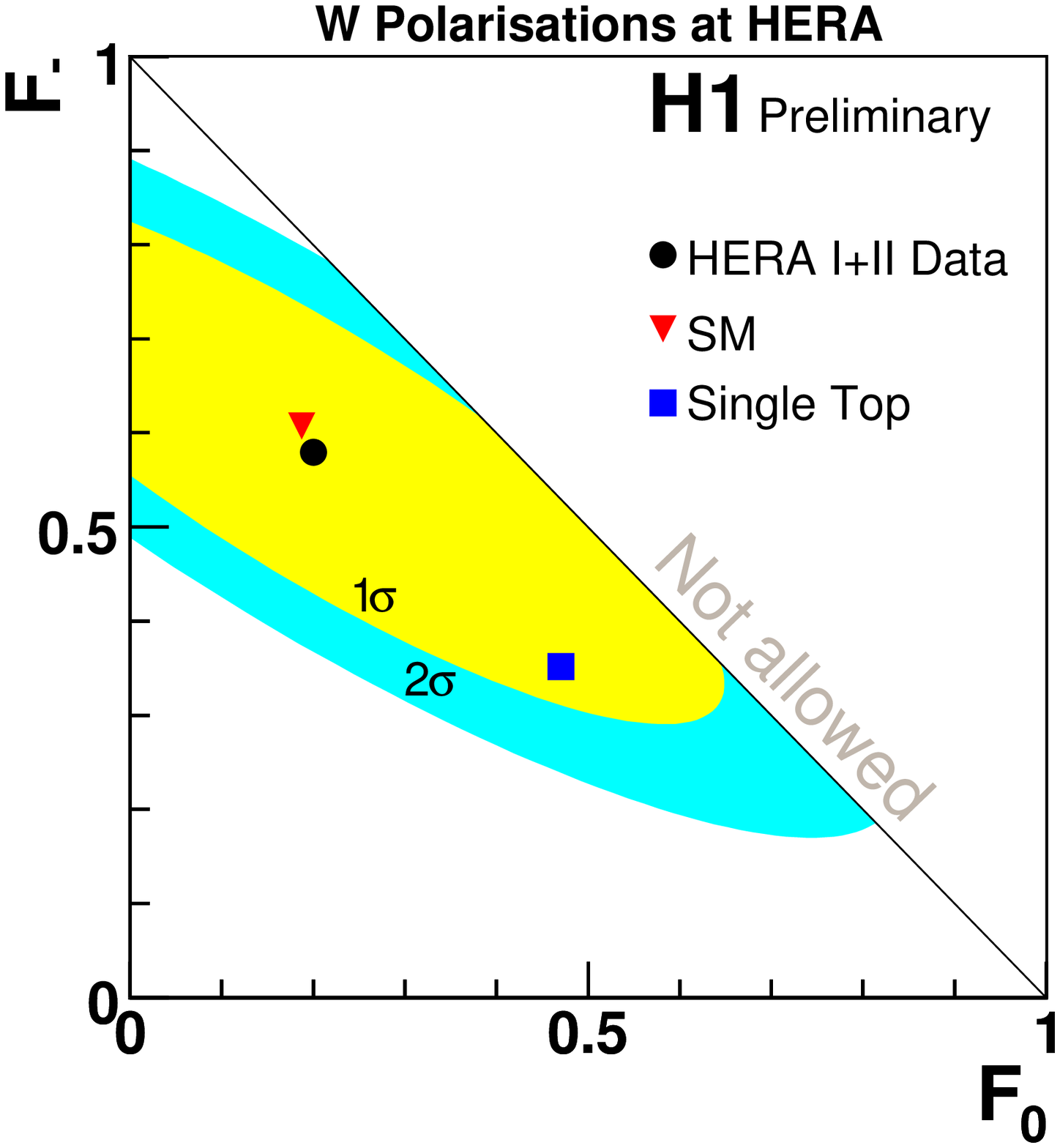}
  \caption{
  	Results from possible extensions to the search for single top production.
  	(a) The hadronic transverse momentum distribution of
	$\tau + P_T^{\rm miss}$ events in the H1 $e^{\pm}p$ HERA I+II data.
	The data (points) are compared to the SM expectation (open histogram).
	The signal component of the SM expectation is given by the
	striped histogram. $\rm N_{Data}$ is the total number of data
	events observed and $\rm N_{SM}$ is the total SM expectation. The
	total uncertainty on the SM expectation is given by the shaded band.	
	%Top: A simultaneous fit (solid histogram) of $F_{-}$ and $F_{0}$ 
	%to the measured differential cross section
	% $d\sigma/d\left(\qcosths\right)$
	%(points), where the error bars denote the statistical uncertainty only.
	%The SM prediction (dashed histogram) is shown with a 15\% theoretical
	%systematic uncertainty (hatched area). Bottom: 
	(b) The fit result for the simultaneously extracted $W$ boson polarisation
	fractions $F_{-}$ and $F_{0}$ (point) with 1 and 2$\sigma$ CL contours. The predictions for
	the SM (triangle) and ANOTOP (square) are shown.}
  \label{fig:tauandpol}
\end{figure}

Limits on the signal cross section are extracted from the discriminator spectra using a
maximum likelihood method~\cite{h1topnew}.
Likelihood functions are calculated for the electron and muon channel separately.
An upper bound on the cross section of $\sigma_{ep \rightarrow e t X}<$~0.16~pb at 95\% CL
is found, which is translated into an upper bound on the coupling $\kappa_{tu\gamma}<$~0.14.

Figure \ref{fig:toplimit} shows existing limits on the anomalous couplings $\kappa_{tu\gamma}$
and $v_{tuZ}$.
Anomalous couplings $\kappa_{tc\gamma}$ of the charm quark are neglected.
The top mass is assumed to be $m_t = 175$~GeV in order to compare with previous results.
Also shown in fig.~\ref{fig:toplimit} are results from the L3 experiment at LEP~\cite{l3top},
the CDF experiment at the Tevatron~\cite{cdftop} using Run~I data and results from the ZEUS 
experiment using HERA~I data~\cite{zeustop}.
A new result from CDF~\cite{cdftopnew} (not shown) derives a limit
on the branching ratio $B(t \rightarrow Zq)$ of $3.7\%$, which translates as an upper limit on
the anomalous vector coupling of $v_{tuZ}\lesssim 0.2$ and is the strictest limit to date.
The preliminary H1 result presented here extends the bound on $\kappa_{tu\gamma}$ into a region so far uncovered by current~colliders.

\section{Extensions}

There are further analyses possible at HERA that have potential to give some information about
anomalous single top production at HERA.

\label{sec:tau}

The H1 Collaboration has also performed a search for events with an isolated tau
lepton and large missing transverse momentum, using the full HERA~I+II $e^{\pm}p$
data~\cite{h1isotaunew}.
This search is complementary to the electron and muon searches described above,
and provides a test of lepton universality.
In addition, some BSM scenarios favour the third lepton generation, which could
lead to an enhancement of tau lepton production.
The tau identification algorithm
exploits the event signature of hadronic tau decays of a narrow, low track 
multiplicity (1--prong) jet in coincidence with missing transverse momentum.
The hadronic transverse momentum distribution of the final sample is shown in
fig.~\ref{fig:tauandpol}(a), where 20 events are observed in the data compared to a SM
prediction of 19.5~$\pm$~3.2.
The latter is dominated by charged current events and the signal
purity is much lower than in the electron and muon channels, at around 14\%.
For $P_{T}^{X}>$~25~GeV one event is selected in the data, compared to a SM
prediction of $0.99~\pm~0.13$ events. 
Due to this difficult background situation no attempt is made to use
the tau channel in the search for single top production.

H1 has also tested another possible handle on single top production via the measurement
of the $W$ boson polarisation fractions. These are expected to be different in SM {\it versus} BSM processes~\cite{wpoltheory}.
The measurement reconstructs the $W$ boson and makes use of the \cosths\ distributions in the 
decay $W\rightarrow e/\mu+\nu$,  where $\theta^{*}$ is defined as the angle between the $W$ boson momentum in the lab frame and the charged decay lepton in the $W$ boson rest frame.
The measured left handed $F_{-}$ and longitudinal $F_{0}$ $W$ boson polarisation fractions,
shown in fig.~\ref{fig:tauandpol}(b), 
are found to be in good agreement with the SM
prediction but also compatible with anomalous single top production within 1$\sigma$
confidence level (CL)~\cite{h1wpol}.

\section{Conclusion}
Anomalous single top production via FCNC has been investigated at HERA. In a sample corresponding
to about $0.5$~fb$^{-1}$ of integrated luminosity, H1 has set an upper limit on 
$\sigma_{ep\rightarrow etX} < 0.16$~pb at $95\%$ CL, corresponding to an upper bound on the
anomalous magnetic coupling $\kappa_{tu\gamma} < 0.14$. This result extends the reach of
previous analyses. Extensions of this search have been explored and tested 
for their ability to contribute to a look at single top production at HERA.

\end{document}